\documentstyle[preprint,prd,aps]{revtex}
\begin{document}
\textwidth=6.0in
\textheight=8.5in
\renewcommand{\d}{{\rm d}}
\titlepage
\thispagestyle{empty}
\preprint{ITP-SB-93-75}

\title{COHERENT STATES IN NULL-PLANE Q.E.D.}

\author{Anuradha Misra}

\address{Department Of Physics and Institute for Theoretical Physics,\\
State University of New York at Stony Brook, New York
11794.}

\maketitle
\begin{abstract}

Light front field theories are known to have the usual infra-red
divergences of the equal time theories, as well as new `spurious' infra-red
divergences. The former kind of IR divergences are usually treated by giving
a small mass to the gauge particle. An alternative method to
deal with these divergences is to calculate the transition matrix
elements in a coherent state basis. In this paper we present, as
a model calculation, the lowest order correction to the three point vertex in
QED using
a coherent state basis in the light cone formalism. The relevant transition
matrix element is shown to be free of the true IR divergences up to
$O(e^2)$.
\end{abstract}
\pacs{ }

\date{\today}
\setcounter{page}{1}
\vspace*{1cm}
\section{INTRODUCTION}
\vspace*{1cm}

Light front field theories have been the subject of considerable interest in
the past
few years as they promise to provide a practical tool for solving the problem
of highly relativistic bound states \cite{PER90,PAU85}. There are two major
approaches to obtaining the bound state wave functions in the light-cone
framework ---the Light Front Tamm Dancoff (LFTD) method \cite{PER90} and the
Discretized Light Cone Quantization (DLCQ) method \cite{PAU85} --- both of
these based on diagonalizing the light cone Hamiltonian in the  basis of Fock
states. Both of these approaches are beset by the usual ultraviolet
divergences of field theory coming from large transverse momenta, as well as
by  infrared divergences near $k^+=0$, where $k^+=k^0+k^1$ is the
longitudinal momentum. The infrared divergences can be classified into two
different categories---`spurious' IR (infra-red) divergences and `true' IR
divergences. The spurious IR divergences are just a manifestation of the
ultraviolet divergences of equal time theory and can be regularized by an IR
cutoff on small values of longitudinal momentum \cite{MUS91,TAN91}.  The `true'
IR divergences are the bona-fide infrared divergences of the equal time theory
and are  present due to particles being on mass shell. These  can be
taken care of by giving the photon a small mass \cite{MUS91,TAN91}.

In the present work we suggest an alternative treatment of this latter kind of
IR divergences. Addressing this excercise may seem unnecessary in the case of
QED, since a simple solution (i.e. giving photon a small mass ) already exists.
However, the suggested formalism may turn out to be useful in future work on
 non-abelian gauge theories, where giving mass to the gauge
particles violates gauge invariance.

Both LFTD and DLCQ are based on the old-fashioned Hamiltonian perturbation
theory, wherein one calculates the matrix elements of the light front
Hamiltonian
between Fock states. In the present paper,  we propose another set of basis
states---the coherent states---to calculate these matrix elements. The
usefulness of coherent states in the proof of cancellation of IR divergences
in equal time theories has been well established \cite{CHU65,KUL70}. Chung has
shown that the IR divergences of QED are eliminated to all orders in
perturbation theory in the matrix elements by an appropriate choice of initial
and final soft photon states\cite{CHU65}. Fadeev and Kulish investigated the
asymptotic behaviour of the QED Hamiltonian and showed that when a particle
mass
becomes negligible compared to the energy scale, the asymptotic Hamiltonian
need
not coincide with the free one. This leads to a redefinition of asymptotic
states\cite{KUL70,DUC89}:

\begin{equation}
\vert{n;\pm}\rangle = \Omega_{\pm}^A \vert{n}\rangle \;,
\end{equation}
where $\Omega_\pm^A$ is the asymptotic evolution operator and $\vert{n}\rangle$
is a Fock state. They further showed the cancellation of IR divergences when
the matrix elements were calculated between these coherent states. In the
Light-Front theories, one can make use of this property of coherent states to
separate the two kinds of IR divergences.

As a first step in this direction, we have, in this
paper, calculated the lowest order radiative correction to the three point
vertex in null plane QED using the coherent state basis. This
calculation has been done in Ref.~\cite{MUS91} --- henceforth referred to as
I---
in the Fock state basis. The main
subjects of I are the ultraviolet and spurious IR divergences, and the
problem of `true' IR divergences has been treated by giving the photon a small
mass. We shall use their expression for the vertex correction {\it without}
giving
the photon a mass, and show that the true IR divergences present in this
expression are cancelled in the coherent state basis by additional
contributions coming from the emission and absorbtion of soft photons.

The present formalism has been developed for continuum light-cone QED.
However, if applicable to discretized case, it may help to provide a
solution to the problem of troublesome zero modes\cite{TAN91}. In discretized
light cone QED, one has to eliminate the $k^+ = 0 $, $k_{\perp} = 0$ state
by an artificial IR cutoff in order to obtain convergence. The present work was
inspired by the hope that the coehrent state formalism developed here may
provide an alternative solution to this problem.

The plan of the paper is as follows: In section II, for the sake of
completeness
we shall set our notation
and present the relevant result of I. In section III, we shall define the
asymptotic
region and coherent states in the light-cone framework.  Section IV presents
 the  main result of this paper. We calculate the transition matrix element
between
coherent states and show the cancellation of true IR divergences. Section V
contains some concluding remarks regarding the possible usefulness of coherent
state formalism in DLCQ calculations. Appendix A contains some useful
properties of coherent states
and Appendix B presents the details of the calculation in section
IV\@.

\vspace*{1cm}
\section{PRELIMINARIES}
\vspace*{1cm}

We shall use the notation of I.

The light-cone QED Hamiltonian is given by \cite{MUS91},
\begin{equation}
P^-= H \equiv H_0 + V_1 + V_2 + V_3 \;,
\end{equation}
where
\begin{equation}
H_0= \int d^2x_\perp dx^- \{ {\imath \over 2} \bar{\xi}\gamma^-
\stackrel{\leftrightarrow} {\partial}_-\xi + {1 \over 2} (F_{12})^2
-{1 \over 2}a_+\partial_- \partial_k a_k \}
\end{equation}
is the free Hamiltonian as a function of independent degrees of freedom, and
\begin{equation}
V_1=e \int d^2 x_\perp dx^- \bar{\xi} \gamma^{\mu}\xi a_\mu\;,
 \end{equation}
is the three point vertex interaction.  $V_2$ and $V_3$ are non-local
effective  four point vertices corresponding to instantaneous fermion and
photon exchange respectively. Detailed expressions for $V_2$ and $V_3$ are not
needed for our purposes, and can be found in I.

$\xi$ and $ a_{\mu}$ can be expanded in terms of creation and annihilation
operators as usual,
\begin{displaymath}
\xi (x) = \int {d^2p_\perp \over {(2 \pi)^{3 \over 2}}}
\int {dp^+ \over {\sqrt{ 2 p^+}}} \sum_{s=\pm {1 \over 2}} [ u(p,s)
e^{-\imath(p^{+}x^- - p_\perp x_\perp)}b(p,s,x^+)
\end{displaymath}
\begin{equation}
\;+ v(p,s)e^{\imath (p^+x^- - p_\perp x_\perp)}d^\dagger
(p,s,x^+)]\;,
\end{equation}
\begin{displaymath}
a_\mu (x) = \int  {d^2q_\perp \over (2 \pi)^{3 \over 2}}
\int {dq^+ \over {\sqrt{2 q^+}}}
\sum_{\lambda=1,2} \epsilon_{\mu}^\lambda(q)
[e^{-\imath(q^+x^- - q_\perp x_\perp)}a(q,\lambda,x^+)
\end{displaymath}
\begin{equation}
\; + e^{\imath(q^+x^- - q_\perp x_\perp)}a^\dagger(q,\lambda,x^+)]\;,
\end{equation}
where
\begin{equation}
\{b(p,s),b^\dagger(p^\prime,s^\prime)\}=
\delta(p^+ - p^{\prime +)}
\delta^2 (p_\perp - p_\perp^\prime) \delta_{s,s^\prime} =
\delta^3 (p - p^\prime) \delta_{s,s^\prime}\;,
\end{equation}
\begin{equation}
\{d(p,s),d^\dagger (p^\prime,s^\prime)\} = \delta(p^+ - p^{\prime +)}
\delta^2 (p_\perp - p_\perp^\prime) \delta_{s,s^\prime} =
\delta^3 (p - p^\prime) \delta_{s,s^\prime}\;,
\end{equation}
\begin{equation}
[a(q,\lambda),a^\dagger (q^\prime,\lambda^\prime)]=
\delta (q^+ - q^{\prime +})
\delta^2 (q_\perp - q_\perp^ \prime) \delta_{\lambda,\lambda^\prime} =
\delta^3 (q - q^\prime) \delta_{\lambda,\lambda^\prime}\;.
\end{equation}

The Hamiltonian can be expressed in terms of annihilation and creation
operators. For example,

\begin{displaymath}
V_1 = e \int d^2x_\perp dx^- \int [dp][d{\bar{p}}]
[dk] \sum_{s,s^\prime,\lambda}
\biggl[e^{\imath \bar{p} \cdot x} \bar{u}(\bar{p},s^\prime)b^\dagger(\bar{p},
s^\prime)
\end{displaymath}
\begin{displaymath}
+ e^{-\imath \bar{p}\cdot x} \bar{v}(\bar{p},s^\prime)d(\bar{p},s^\prime)
\biggr]
 \gamma^\mu \biggl[ e^{-\imath p\cdot x} u(p,s)b(p,s)
\end{displaymath}
\begin{equation}
+ e^{\imath p \cdot x}
v(p,s)d(p,s)^\dagger\biggr] \epsilon ^{\lambda}_{\mu}(k)\biggl[e^{-\imath k
\cdot x}a(k,\lambda)
\; \;+ e^{\imath k \cdot x} a^\dagger(k,\lambda)\biggr]\;,
\label{eq:int} \end{equation}

where

\begin{equation}
\int [dp^\prime] \equiv \int_{-\infty}^{\infty} {d^2p_\perp \over {(2
\pi)^{3 \over 2}}} \int_0^\infty{dp^+ \over {\sqrt{2p^+}}}
\end{equation}

Similar expressions for $ V_2 $ and $ V_3 $ can be obtained and are given in
Appendix A of Ref.\ \cite{MUS91}.

In perturbative  light-cone  QED, all graphs are matrix elements of the
transition matrix T given by \cite{MUS91},

\begin{equation}
T= V + V {1 \over {p^- - H_0}}V + \cdots
\end{equation}
between the Fock states. For example, the lowest order correction to the 3
point vertex ${\Lambda}^{\mu}(p,\bar{p})$ is given by the
following matrix element,

\begin{displaymath}
T_{21}= \epsilon^{\lambda}_{\mu}(q)
\Lambda^{\mu}(p,\bar p)
\end{displaymath}
\begin{displaymath}
=\langle \bar{p},\sigma,q,\lambda \vert V_1 {1 \over {p^- - H_0}}
V_1 {1 \over {p^- - H_0}}V_1 \vert p,s \rangle +
\langle \bar{p},\sigma,q,\lambda \vert V_2{1 \over {p^- - H_0}}V_1
\vert p,s \rangle
\end{displaymath}
\begin{equation}
+ \langle \bar{p},\sigma,q,\lambda \vert V_3{1 \over {p^- - H_0}}V_1
\vert p,s \rangle \;,
\end{equation}
where
$\vert 1 \rangle\ = \vert p,s \rangle$ is the Fock state containing a single
fermion and $\vert 2 \rangle = \vert
\bar{p},\sigma,q,\lambda \rangle$ is the Fock state containing one fermion and
one photon.

The full set of diagrams corresponding to the above matrix element is given in
Ref.\ \cite{MUS91}. We shall limit ourselves to the calculation of
$\Lambda^+(p,\bar{p} ) $, in which case many of the diagrams do not
contribute due to their tensor structure. The only diagrams contributing to
$\Lambda^+(p,\bar{p})$ are shown in Fig.\ \ref{reg}.
Calculation of the diagrams in Fig.\ \ref{reg} has been done in I. Later we
shall make one further simplification, i.e. we shall consider only $q=0$ case.
In
this case, only diagram in Fig.\ \ref{reg}(a) contributes to the matrix element
in question\cite{MUS91}.

One can calculate the diagram in Fig.\ \ref{reg}(a) from
\begin{equation}
T_{21}=
\langle p^{\prime},\sigma,q,\lambda \vert V_1 {1 \over {p^- - H_0}}
V_1 {1 \over {p^- - H_0}}V_1 \vert p,s \rangle
\end{equation}
by substituting $V_1$ from Eq.~(\ref{eq:int}). Using the  relations
\begin{equation}
d_{\mu\nu}(p) = \sum_{\lambda=1,2} \epsilon^\lambda_\mu(p)\epsilon^\lambda_\nu
(p)
=-g_{\mu\nu} + {{\delta_{\mu+}p_\nu+\delta_{\nu+}p_\mu} \over p^+}\; ,
\label{eq:pot}
\end{equation}
and
\begin{equation}
\sum_{s=\pm 1/2} u(p,s)\bar{u}(p,s)=\not{p} + m \;,
\label{eq:uubar}
\end{equation}
 one obtains, after
 a
straightforward calculation \cite{MUS91},
\begin{equation}
\Lambda^\mu_a(p,\bar{p})=\lambda e^3 \int {{d^2 k_\perp} \over {(4 \pi)^3}}
\int {{dk^+} \over {k^+k^{\prime +}k^{\prime \prime +}}} {{N_a^\mu + yN_b^\mu}
\over {(p^- -k^- -k^{\prime -})(p^- -k^- -k^{\prime \prime -}-q^-)}} \;,
\end{equation}
where
\begin{equation}
N^{\mu}_a + yN^{\mu}_b = \bar{u} (\bar{p}, \sigma)\gamma^a(\not{k}^{\prime}+m)
\gamma^{\beta}d_{\alpha \beta}
(\not{k}^{\prime \prime}+m)\gamma^\mu  u(p,s) \;,
\end{equation}
and
\begin{equation}
\lambda= (2\pi)^{3 \over 2}\sqrt{2p^+}\sqrt{2{\bar{p}}^+}\sqrt{2q^+} \;.
\end{equation}
Reparametrizing the momentum variables as
\begin{equation}
k=(xp^+,{{(xp_{\perp} + k_{\perp})^2} \over {2p^+x}},xp_\perp+k_\perp)\;,
\end{equation}
\begin{equation}
q=y(p^+,{{p^2_{\perp}} \over {2p^+}},p_\perp)\;,
\end{equation}
\begin{equation}
k^{\prime}=((1-x)p^+,{{((1-x)p_{\perp} - k_{\perp})^2 + m^2} \over
{2(1-x)p^+}},(1-x)p_\perp - k_\perp) \;,
\end{equation}
\begin{equation}
k^{\prime \prime}=((1-x-y)p^+,{{((1-x-y)p_{\perp} - k_{\perp})^2 + m^2} \over
{2(1-x-y)p^+}},(1-x-y)p_\perp - k_\perp)\;,
\end{equation}
\begin{equation}
\bar{p}=((1-y)p^+,(1-y){{p_\perp^2 } \over {2p^+}}+{m^2 \over {2p^+(1-y)}},
(1-y)p_\perp) \;,
\end{equation}
and using the properties of $\gamma$-matrices, one finally obtains after some
algebra,
\begin{displaymath}
\Lambda^+_a (p,\bar{p}) = \lambda e^3 \int_\alpha^{1-y} dx \int {{d^2k_\perp}
\over
{(2\pi)^3}} {{\bar{u} (\bar{p},\sigma)\gamma^+ \bigl({{1-x} \over x} \bigr)
\bigl[{1 \over {1-x}} + 1-x \bigr] u(p,s)} \over  {(k_\perp^2 + m^2 x^2)}}
\end{displaymath}
\begin{equation}
+ \lambda e^3 \int_\alpha^1 dx \int {{d^2k_\perp} \over
{(2\pi)^3}} {{\bar{u} (\bar{p},\sigma)2x(1-x)[(1-x)\not{p}p^+ - (1-x)m^2
\gamma^+ - p^+m]u(p,s)} \over  {[k_\perp^2 + m^2 x^2][k_\perp^2 + g_1
m^2)]}}\;,
\end{equation}
where
\begin{equation}
g_1 ={{x(x+y)} \over {(1-y)}} \;.
\end{equation}
Putting $q=0$ and ignoring  the IR convergent contribution to $\Lambda^+_a $,
one finally arrives at
\begin{equation}
\Lambda^+_{IR}(p,p) = 4 \lambda p^+ e^3 \int_\alpha^1{dx \over x}
\int {{d^2k_\perp} \over
{(2\pi)^3}} {1 \over {k_\perp^2 + m^2x^2}} -
4 \lambda p^+ e^3 \int_\alpha^1 dx \int {{d^2k_\perp} \over
{(2\pi)^3}} {{m^2x} \over {\bigl(k_\perp^2 + m^2x^2 \bigr)^2}}\;,
\label{eq:irdiv}
\end{equation}
where $\alpha$ is an infrared regulator, which eliminates the IR divergences
near  $k^+ = 0$. The first term in the above expression has ultraviolet
divergences also, which are usually regularized by dimensionanl regularization
or by putting a cutoff on large values of $k_\perp$.

\vspace*{1cm}
\section{INFRARED DIVERGENCES AND THE COHERENT STATE BASIS}
\vspace*{1cm}

 For
massive particles as well as for massless particles with  $k_\perp \ne 0$,
the condition $k^+>\alpha$  as in Eq.~(\ref{eq:irdiv})   is equivalent to
putting an ultraviolet regulator
on large values of $k_3$ in the usual space time formulation. However, for a
massless particle at $k_\perp=0$, the divergences near $k^+=0$ are the true IR
divergences of equal time theory, and in this region the condition $k^+>\alpha
$
does not
follow from the condition of an ultraviolet cutoff on $k_3$.
The IR divergences in equal time QED are cancelled in the cross sections
when a sum is taken over all possible initial and final states with any number
of soft photons having momenta below the threshold of obesrvability.
Chung \cite{CHU65} suggested that the origin of IR divergences lies  in an
inappropriate choice of initial and final states to represent the experimental
situation and showed that the matrix elements do not have IR divergences if
initial and final states are chosen to be appropriately defined coherent states
instead of the usual Fock states. Kulish and Faddev \cite{KUL70}  argued that
since the asymptotic Hamiltonian does not coincide with the free one in QED,
the matrix elements should be calculated between coherent states instead of
the Fock states. They obtained a form  for the asymptotic states starting from
the asymptotic Hamiltonian. In the following, we
shall obtain the form of coherent states in the light-cone formalism following
the same procedure. In this way, we will extend the coherent state formalism
to the light-cone field theory.
The light-cone time dependence of the interaction Hamiltonian is given by
\begin{equation}
H_I(x^+)= e\sum_{i=1}^4 \int d\nu_i[ e^{-\imath \nu_i x^+} {\tilde h}_i(\nu_i)
+ e^{\imath \nu_i x^+}
{\tilde h}^ {\dagger}_i (\nu_i)]
\end{equation}
where $ { \tilde h}_i(\nu_i)$ are  the QED interaction vertices:
\begin{equation}
{\tilde h}_1 = \sum_{s,s^{\prime},\lambda} b^{\dagger}(\bar p,s^\prime)b(p,s)
a(k,\lambda) \bar u(\bar
p,s^\prime)\gamma^{\mu}u(p,s)\epsilon^{\lambda}_{\mu}\;,
\end{equation}
\begin{equation}
{\tilde h}_2 = \sum_{s,s^{\prime},\lambda} b^{\dagger}(\bar p,s^\prime)
d^{\dagger}(p,s)
a(k,\lambda) \bar u(\bar
p,s^\prime)\gamma^{\mu}v(p,s)\epsilon^{\lambda}_{\mu}\;,
\end{equation}
\begin{equation}
{\tilde h}_3 = \sum_{s,s^{\prime},\lambda} d(\bar p,s^\prime)b(p,s)
a(k,\lambda) \bar v(\bar
p,s^\prime)\gamma^{\mu}u(p,s)\epsilon^{\lambda}_{\mu}\;,
\end{equation}
\begin{equation}
{\tilde h}_4 = \sum_{s,s^{\prime},\lambda} d^{\dagger}(\bar p,s^\prime)d(p,s)
a(k,\lambda) \bar v(\bar
p,s^\prime)\gamma^{\mu}v(p,s)\epsilon^{\lambda}_{\mu}\;,
\end{equation}
 and
$\nu_i$ is  the light cone energy transferred at the vertex ${\tilde h}_i$ .
The integration measure is given by
\begin{equation}
\int d\nu =
 {1 \over{{(2 \pi)}^{3/2}}} \int {{[dp][dk]}\over{\sqrt{2\bar p^+}}} \;,
\end{equation}
$\bar p^+$ and $\bar p_{\perp}$ being fixed at each vertex by momentum
conservation. For example,
\begin{equation}
\nu_1 = p^- + k^- - \bar p^- = {p \cdot k \over p^++k^+}
\end{equation}
is the energy transfer at $ee\gamma$ vertex.

At asymptotic limits, non-zero contributions to $H_I(x^+)$ come from regions
where $\nu_i$ goes to zero. It is easy to see that $\nu_2$ and $\nu_3$  are
always non-zero, and hence, ${\tilde h}_2$ and ${\tilde h}_3$ do not appear in
the asymptotic Hamiltonian. Thus, the asymptotic Hamiltonian is defined by the
following expression,
\begin{equation}
V_{as}(x^+) = e \sum_{i=1,4} \int d\nu_i \Theta_\Delta(k)
 [ e^{-\imath \nu_i x^+} \tilde {h_i}(\nu_i)
+ e^{\imath \nu_i x^+}
\tilde h^ {\dagger}_i (\nu_i)] \;,
\label{eq:asmh}
\end{equation}
where $\Theta_{\Delta}(k)$ is a function which takes a value 1 in the
asymptotic region and is zero elsewhere.

 One can define the asymptotic region to consist of all points in the phase
space for which
\begin{equation}
{{p \cdot k} \over p^+} < \Delta E     \;,
\end{equation}
where $\Delta E$ is an energy cutoff which may be chosen to be the experimental
resolution. For simplicity, we shall choose a frame $p_\perp = 0 $. In this
frame the above condition reduces to
\begin{equation}
{{p^+k_\perp^2} \over 2k^+} + {m^2k^+ \over 2p^+} < \Delta \;,
\label{eq:asrn}
\end{equation}
where  $\Delta = p^+\Delta E$.

Thus, for all the points satisfying Eq.~(\ref{eq:asrn}), $\nu_1$ and $\nu_4$
can be
approximated by zero. This implies that in  this region, the  asymptotic
Hamiltonian is different from the free Hamiltonian. For the present purposes,
i.e. in order to eliminate the true IR divergences, we find it sufficient to
choose a subregion of the above mentioned region as the asymptotic
region. We define this subregion to be consisting of all points
$(k^+, k_\perp)$
satisfying:
\begin{equation}
k_\perp ^2 < {{k^+ \Delta} \over{p^+}}  \;,
\end{equation}
\begin{equation}
k^+ < {{p^+ \Delta} \over {m^2}}   \;.
\end{equation}
This choice of the asymptotic region leads to the asymptotic
interaction Hamiltonian defined by Eq.~(\ref{eq:asmh}) with

\begin{equation}
 \Theta_\Delta(k)=\theta({{k^+\Delta} \over p^+} - k_\perp^2)
\theta({{p^+\Delta} \over m^2} - k^+)
\label{eq:theta1}
\end{equation}

The asymptotic states can be defined in the usual manner by \cite{DUC89},
\begin{equation}
\vert n \colon coh \rangle = \Omega_{\pm}^A \vert{n}\rangle    \;,
\end{equation}
Xwhere $\vert{n}\rangle$ is a Fock state of charged particles and hard photons
and $\Omega_{\pm}^A$ are the asymptotic M\"oller operators defined by
\begin{equation}
\Omega_{\pm}^A = T\biggl[ -\imath \int^0_{\mp} V_{as}(x^+)dx^+ \biggr] \;.
\end{equation}
Following the standard procedure \cite{KUL70} of substituting $k^+=0$,
$k_{\perp}=0$ in all
the slowly varying functions of $k$,  and  carrying out the $x^+$
integration, we
arrive at the following expression for the asymptotic states:
\begin{displaymath}
\Omega_{\pm}^A \vert n \colon p_i \rangle =
exp\biggl[-e\int {{[dp]} \over {\sqrt{2{\bar p}^+}}} \int \sum_{\lambda=1,2}
 {{dk^+} \over
{\sqrt{2k^+}}} \int {{d^2k_\perp} \over {(2\pi)^{3 \over 2}}}
\end{displaymath}
\begin{equation}
[f(k,\lambda,p) a^\dagger(k,\lambda) - f^*(k,\lambda,p)a(k,\lambda)]\rho(p)
\biggr]
\vert n \colon p_i \rangle \;,
\end{equation}
where
\begin{equation}
f(k,\lambda \colon p) = {{p_\mu\epsilon_\lambda^\mu(k)} \over {p\cdot k}}
\theta({{k^+\Delta} \over p^+}-k_\perp^2)
\theta({{p^+\Delta} \over m^2}-k^+) \;,
\label{eq:theta2}
\end{equation}

\begin{equation}
f(k,\lambda \colon p) = f^*(k, \lambda \colon p) \;,
\end{equation}
if one follows the convention in I for photon polarization and
\begin{equation}
\rho(p) = \sum_n\bigl[b_n^\dagger(p)b_n(p) - d_n^\dagger(p)d_n(p) \bigr]\;.
\end{equation}
Applying the operator $\rho(p)$ on  the Fock state, we finally obtain,
\begin{equation}
\Omega_{\pm}^A \vert n \colon p \rangle =
exp\biggl[-e\int \sum_{\lambda=1,2}
 {{dk^+} \over
{\sqrt{2k^+}}} \int {{d^2k_\perp} \over {(2\pi)^{3 \over 2}}}[f(k,
\lambda \colon p) a^\dagger(k,\lambda) - f^*(k,\lambda \colon p)a(k,\lambda)]
\biggr]
\vert n \colon p \rangle \;.
\end{equation}
In particular, the coherent state containing one fermion is given by
\begin{equation}
\vert p,\sigma \colon f(p) \rangle = exp\biggl[-e\int \sum_{\lambda=1,2}
 {{dk^+} \over
{\sqrt{2k^+}}} \int {{d^2k_\perp} \over {(2\pi)^{3 \over 2}}}[f(k,
\lambda \colon p) a^\dagger(k,\lambda) - f^*(k,\lambda \colon p)a(k,\lambda)]
\biggr]  \vert
p,\sigma \rangle \;,
\label{eq:state1}
\end{equation}
and the coherent state containing one fermion and one hard photon is given by
\begin{equation}
\vert p,\sigma,q,\lambda \colon f(p) \rangle = exp\biggl[-e\int
\sum_{\lambda=1,2} {{dk^+} \over {\sqrt{2k^+}}} \int {{d^2k_\perp} \over
{(2\pi)^{3 \over 2}}}\bigl[f(k,\lambda \colon p)
a^\dagger(k,\lambda) - f^*(k,\lambda \colon p)a(k,\lambda)\bigr] \biggr]
\vert p,\sigma,q,\lambda \rangle\;.
\label{eq:state2}
\end{equation}

Some useful properties of these coherent states are listed in Appendix B.

\vspace*{1cm}
\section{CALCULATION OF VERTEX CORRECTION IN COHERENT SPACE BASIS}
\vspace*{1cm}

Let us rewrite the IR divergent contribution to the $ O(e^2) $ vertex
correction
as given by Eq.~(\ref{eq:irdiv}) as
\begin{equation}
\Lambda^+_{IR}(p,p) = \int_0^1 dx \int d^2k_\perp I(x,k_\perp)  \;.
\end{equation}
{}From our discussion in the previous section, it is natural to split the $x$
and
$k_{\perp}$ integrals in the above equation
as
\begin{displaymath}
\Lambda^+_{IR}(p,p) = \int_0^1 dx \int d^2k_\perp I(x,k_\perp)
\theta({{k^+\Delta} \over p^+}-k_\perp^2)\theta({{p^+\Delta} \over m^2}-k^+)
\end{displaymath}
\begin{displaymath}
+ \int_{{\Delta} \over {m^2}}^1
dx \int {d^2k_\perp} I(x,k_\perp)
\end{displaymath}
\begin{equation}
+  \int^{{\Delta } \over {m^2}}_\alpha dx \int {d^2k_\perp} I(x,k_\perp) \theta
(k_\perp^2 -{{k^+\Delta} \over {p^+}}) \;.
\end{equation}
In what follows, we will show that the vertex correction does not have the
true
IR divergences like the first term in the above equation
if one calculates the matrix elements of $T$ between
coherent states defined by Eqs.~(\ref{eq:state1}) and (\ref{eq:state2}).

In the basis of coherent states there are additional $ O(e^3)$
contributions to $T_{21}$ coming from
\begin{eqnarray}
T_{21}^\prime &=
\langle \bar{p},\sigma,q,\gamma \colon f(\bar{p}) \vert V_1 {1 \over {p^- -
H_0}}V_1\vert p,s \colon f(p) \rangle \nonumber \\
&=\epsilon^\lambda_\mu(q)\Lambda^{\prime\mu}(p,\bar{p}) \;,
\end{eqnarray}
corresponding to the emission and absorbtion of soft photons, as shown in
Fig.\ref{coh}.
Details of the calculation of diagrams in Figs.~\ref{coh}(a), (b) and (c)
are given
in Appendix B. Here we will
present the final expressions for $\Lambda^+_{2a}$, $\Lambda^+_{2b}$ and
$\Lambda^+_{2c}$, first,
\begin{equation}
\Lambda^+_{2a} =-{\lambda e^3 \over {8\pi^3}} \int {dk^+ \over k^{+2}}
\int d^2k_\perp \Theta_\Delta(k)\bigl({{p^+-k^+} \over {p \cdot k}} \bigr)
\bigl[1 + {{p^+k^- - p^-k^+} \over k \cdot p} \bigr] \;.
\end{equation}
For the case $p_\perp = 0$, the above equation reduces to
\begin{equation}
\Lambda^+_{2a}=-{{\lambda e^3} \over {\pi^3}} \int {dk^+
\over 2k^+}(p^+ - k^+)
\int d^2k_\perp
{{\Theta_{\Delta}(k)} \over {k_\perp^2 + {m^2k^{+2} \over p^{+2}}}} \biggl[
1-{{m^2k^{+2} \over p^{+2}} \over {k_\perp^2 + {m^2k^{+2} \over p^+2}}} \biggr]
\;.
\end{equation}
Similarly, we find
\begin{equation}
\Lambda^+_{2b} = \Lambda^+_{2c} =
-{{\lambda e^3} \over {4\pi^3}} \int {dk^+ \over k^+} \int d^2k_
\perp\Theta_
\Delta(k){1 \over {p^- -
\bar{p}^-}}\biggl[1-{{p^-k^+} \over {k \cdot p}}\biggr] \;.
\end{equation}
Both $\Lambda^+_{2b}$ and $\Lambda^+_{2c}$ have a vanishing denominator at
$q=0$,
and
can be evaluated using Heitler method \cite{MUS91,HEIT54}, to give
\begin{equation}
\Lambda^+_{2b} + \Lambda^+_{2c}
=-{{\lambda e^3} \over {2\pi^2}} \int dk^+ \int d^2k_\perp \bigl[{k^2_\perp
\over {k_
\perp^2 + {m^2k^{+2} \over
p^+2}}}\bigr]\Theta_\Delta (k) \;,
\end{equation}
which does not have any IR divergence.
So, the  IR divergent contribution to $\Lambda^{\prime+}(p,\bar{p})$ at $q=o$,
 comes
from
Fig.\ \ref{coh}(a)  only,
\begin{equation}
\Lambda^{\prime +}(p,p) = - {{\lambda e^3 p^+} \over {2\pi^3}} \int {dk^+
\over
k^+} \int d^2k_\perp {\Theta_\Delta(k) \over {\bigl[k_\perp^2
+ {m^2(k^+)^2 \over
(p^+)^2}\bigr]}}
\biggl[1 -  {{m^2k^{+2} \over p^{+2}} \over {(k_\perp^2 + {m^2(k^+)^2 \over
(p^+)^2}}}
\biggr] \;.
\label{eq:irdiv1}
\end{equation}

Comparing Eq.~(\ref{eq:irdiv})  and Eq.~(\ref{eq:irdiv1}) one can easily
see that the true
IR divergences
in Fig.\ \ref{reg} and in Fig.\ \ref{coh} cancel exactly. This completes the
proof
of cancellation of  `true' IR divergences in $O(e^2)$ 3-point vertex
correction in the coherent state basis.

\vspace*{1cm}
\section{CONCLUSION}
\vspace*{1cm}

We have presented a lowest order calculation in continuum light-cone QED to
show the cancellation of true IR divergences when a coherent state basis is
used to calculate the matrix elements.

The present calculation has been done in
 continuum case, but the suggested method of using the coherent states as the
asymptotic states in order to calculate the Hamiltonian matrix elements
promises to be useful in  discrete case as well. In particular, it may
be relevant to the problem of  zero modes in
discretized light-cone QED \cite{TAN91,KRAUT92}. Tang  has shown
that in a numerical
DLCQ calculation of energy levels of positronium, the lowest energy level
diverges with K, the harmonic resolution, if one does not remove the
$k^+=0$, $k_{\perp}=0$ state by an artificial IR cutoff. Kr\"autgartner {\it
et\ al}. have
analyzed the various approximations to the DLCQ matrix equation for positronium
and have discussed the singularity occuring due to the exchange of a zero mode
photon. They have claimed that even though one can remove the true IR
divergences by eliminating the $k^+=0$,
$k_{\perp}=0$ state by an artificial cutoff or by giving a small mass to the
photon, yet neither of these procedures leads to convergent results. In order
to achieve convergence, one has to add and subtract an appropriate term to the
Hamiltonian. This counterterm removes the discretized IR divergence and
replaces the term at small $k^+$ and $k_{\perp}$ by the appropriate continuum
value. However, if one calculates the Hamiltonian matrix elements between the
coherent states {\it before}
 discretization is carried out, one may be able to remove the true IR
divergences
as well as reproduce the extra term needed for convergence in a natural manner.
Work in this direction is in progress.

\vspace*{1cm}
\acknowledgements
\vspace*{1cm}

It gives me  great pleasure to acknowledge Prof.\ George  Sterman for
introducing me to the subject. But for his help and encouragement and for many
useful suggestions, this work would not have been possible. I would also like
to thank Prof.\ Gene D.\ Sprouse, Chair, Department of Physics, State
University of New York at Stony Brook for financial support.

\vspace*{1cm}
\centerline{\bf Appendix A}
\vspace*{1cm}

{\it Properties of Coherent States}

We denote with $\vert 1 \colon p_i \rangle$ the coherent state containing a
fermion and a
 superposition of
infinite number of soft photons as defined by Eq.~(\ref{eq:state1}) and with
$\vert 2 \colon p_{\imath}, k_{\imath} \rangle $ the
 coherent state containing a fermion and a hard photon as defined by
Eq.~(\ref{eq:state2}).

It can be  shown easily that the coherent states $\vert 1\colon p_i\rangle$
 are the eigenstates of $a(k,\lambda)$,
\begin{equation}
a(k,\rho)\vert 1 \colon p_i \rangle =
-{e \over {(2\pi)^{3/2}}}{1 \over \sqrt{2k^+}}
f(k,\rho \colon p_i) \vert 1 \colon p_i\rangle \;.
\label{eq:eigen1}
\end{equation}

Also,
\begin{equation}
a(k,\rho) \vert 2\colon p_i,k_i\rangle =
-{e \over {(2\pi)^{3/2}}}{1 \over \sqrt{2k^+}}
f(k,\rho \colon p_i) \vert 2 \colon p_i, k_i \rangle +
\delta^3(k-k_i)\delta_{\rho{\lambda_i}} \vert 1\colon p_i \rangle
\label{eq:eigen2} \;,
\end{equation}
and
\begin{equation}
a^{\dagger}(k,\rho) \vert 1\colon p_i \rangle =
-{e \over {(2\pi)^{3/2}}}{1 \over \sqrt{2k^+}}
f^*(k, \rho:p_i) \vert 1\colon p_i \rangle + \vert 2 \colon p_i, k_i \rangle
\;.
\label{eq:eigen3}
\end{equation}
In the lowest order, , Eq.~(\ref{eq:eigen3}) reduces to
\begin{equation}
a^{\dagger}(k, \rho) \vert 1 \colon p_i \rangle = \vert 2 \colon p_i, k_i
\rangle \;.
\label{eq:eigen4}
\end{equation}
Coherent states satisfy the following orthonormalization properties
\begin{displaymath}
\langle 1\colon p_f,\sigma_f \vert 1 \colon p_i,\sigma_i\rangle =
\delta^{(3)}(p_i-p_f)\delta_{\sigma_i\sigma_f}
\end{displaymath}
\begin{equation}
\langle 2 \colon p_f,\sigma_f, k_f,\lambda_f \vert 1\colon p_i,\sigma_i\rangle
 = \delta^{(3)}(p_i-p_f)\delta_{\sigma_i\sigma_f}f(k_f,\lambda_f\colon p_f)
{e \over (2\pi)^{3/2}}
\end{equation}

\vspace*{1cm}
\centerline{\bf Appendix B}
\vspace*{1cm}

In this appendix, we shall calculate the various contibutions to the matrix
element considered in section IV :

\begin{displaymath}
T^\prime_{21}=
\langle \bar{p},\sigma,q,\lambda \colon f(\bar{p})\vert V_1 {1 \over
{p^- - H_0}}V_1  \vert p,s \colon f(p) \rangle
\end{displaymath}
Substituting for $V_1$ from Eq.~(\ref{eq:int}) and using Eqs.~(\ref{eq:eigen1})
and  (\ref{eq:eigen2}) one obtains
\begin{equation}
T_{21}^\prime = T_{2a} + T_{2b} +T_{2c} \;,
\end{equation}
where $T_{2a}$ and $T_{2b}$  correspond to Feynman diagrams in
Figs.~\ref{coh}(a) and (b)
representing the  absorbtion of soft photons in the initial and
final states and and $T_{2c}$ corresponds to the emission of soft photons from
the final state.
There is no diagram corresponding to the emission of soft photons in the
initial state to this order
as evident from Eq.~(\ref{eq:eigen4}). $T_{2a}$, $T_{2b}$ and $T_{2c}$ are
defined by
\begin{displaymath}
T_{2a}= \langle \bar{p},\sigma  \colon f(\bar{p}) \vert V_a^\dagger {1 \over
{p^- - H_0}} V_a \vert p,s \colon f(p)\rangle \;,
\end{displaymath}
\begin{equation}
T_{2b} =\langle \bar{p},\sigma \colon f(\bar{p}) \vert V_a^\dagger {1 \over
{p^- - H_0}}
V_a^\dagger \vert p,s \colon f(p)\rangle \;,
\label{eq:tmb}
\end{equation}
and
\begin{equation}
T_{2c} = \langle \bar{p},\sigma \colon f(\bar{p}) \vert V_a {1 \over
{p^- - H_0}}
V_a^\dagger \vert p,s \colon f(p)\rangle \;,
\label{eq:tmb1}
\end{equation}
where
\begin{displaymath}
V_a = e \int d^2x_\perp dx^- \int [dp][d{\bar{p}}]
[dk] \sum_{s,s^\prime,\lambda}
[e^{\imath(\bar{p}-p-k) \cdot x}\bar{u}(\bar{p},s^\prime)\gamma^\mu
\end{displaymath}
\begin{equation}
u(p,s)b^\dagger(\bar{p},s^\prime)b(p,s)a(k,\lambda)\epsilon ^{\lambda}_{\mu}
(k)]
\label{eq:tma}
\end{equation}
Using Eqs.~(\ref{eq:uubar}),(\ref{eq:eigen1}) and (\ref{eq:tma}) one obtains in
a straightforward manner
\begin{displaymath}
T_{2a}= \epsilon^\lambda_\mu(q)\Lambda_{2a}^\mu(p,\bar{p})
\end{displaymath}
\begin{displaymath}
\; \; =-{{\lambda e^3 \epsilon_\mu^\lambda(q)} \over 2p^+} \int
{{dk^+d^2k_\perp} \over (2\pi)^3 2k^+} {1 \over {k^- + (p-k)^- - p^-}}
\end{displaymath}
\begin{equation}
\times \sum_{\lambda_1} \epsilon_\nu^{\lambda_1}(k)f(k,\lambda_1 \colon p)
\bigl[\bar{u}(\bar{p},\sigma)\gamma^\mu(\not{p}+m)\gamma^\nu u(p-k,s) \bigr]
\;,
\end{equation}

Alternatively,
\begin{displaymath}
\Lambda_{2a}^\mu(p,\bar{p})=  {{\lambda e^3 } \over 2p^+} \int
{dk^+ \over 2k^+} \int {d^2k_\perp \over (2\pi)^3 } {1 \over {p^- - k^- - (p-k)
^-
}}
\end{displaymath}
\begin{equation}
\times \bigl[\bar{u}(\bar{p},\sigma)\gamma^\mu(\not{p}+m)\gamma^\nu u(p,s)
\bigr]
\sum_\lambda \epsilon_\nu^\lambda (k) \epsilon_{\rho\lambda} (k) f^\rho(k,p)
\;,
\label{eq:lamb3a}\end{equation}
where we have used Eqs.~(\ref{eq:theta1}) and  (\ref{eq:theta2})) in the above
expression, and have approximated
$u(p-k,s)$ by $u(p,s)$. We have defined $f^\rho(k,p)$ by
\begin{equation}
f(k,\lambda \colon p) = f^\rho(k, p)\epsilon^\lambda_\rho(p)\;.
\end{equation}
For $\mu=+$, Eq.~(\ref{eq:lamb3a}) reduces to
\begin{displaymath}
\Lambda^+_{2a}(p,\bar{p})={{\lambda e^3} \over {8\pi^3}} {1 \over 2p^+}
\int {dk^+ \over k^{+2}}\int d^2k_\perp
\end{displaymath}
\begin{equation}
\biggl[{{ \bigl( \bar{u}(\bar{p},\sigma)\gamma^+ (\not p + m) \gamma^\nu u(p,s)
\bigr)\bigl(-f_\nu(k,p)+{k_\nu \over k^+}f^+(k,p)+{\delta_{\nu+} \over k^+}k
\cdot f\bigr)} \over {p^- -k^- - (p-k)^-}}\biggr] \;.
\label{eq:lamb3a1}
\end{equation}

In a frame where $p_\perp = 0 $, one can use Eqs.~(\ref{eq:theta2}) and the
following relations
\cite{MUS91},
\begin{equation}
p^- - k^- -(p-k)^- = - {{p \cdot k} \over {p^+-k^+}}
\end{equation}
\begin{equation}
p \cdot k = {{p^+} \over {2k^+}} \bigl[k_\perp^2 +{{m^2k^2} \over {p^{+2}}}
\bigr]
\end{equation}
to reduce Eq.~(\ref{eq:lamb3a1}) at $q=0$ to

\begin{displaymath}
\Lambda^+_{2a}(p,\bar{p})=-{{\lambda e^3} \over {4\pi^3}}
\int {dk^+ \over k^{+2}}\int d^2k_\perp \theta({{\Delta p^{+}} \over m^2}-k^+)
\end{displaymath}
\begin{equation}
\times \theta({k^+\Delta \over p^+}-k_\perp^2){{p^+ - k^+} \over {p^+}}
\bigl[ 1 + {{p^+k^-
- p^-k^+} \over {k \cdot p}}\bigr] \;,
\end{equation}
which has an IR divergent part given by
\begin{displaymath}
\Lambda^+_{2aIR}(p,p)
=-{{\lambda e^3} \over {2\pi^3}} \int {dk^+ \over k^{+}}\int d^2k_\perp
\theta({{\Delta p^{+}} \over m^2}-k^+)
\end{displaymath}
\begin{equation}
\times \theta({k^+\Delta \over p^+}-k_\perp^2)\bigl[ {{1} \over {(k_\perp^2
+{m^2k^{+2} \over p^{+2}}})} \bigr] \biggl[ 1 - {{m^2k{+2} \over p^{+2}} \over
{k_\perp^2 +{m^2k^{+2} \over p^{+2}}}} \biggr] \;.
\end{equation}

Similarly, one can show that
\begin{displaymath}
\Lambda^+_{2bIR}(p,\bar{p}) = \Lambda^+_{2cIR}(p,\bar{p})
=-{{\lambda e^3} \over {4\pi^3}} \int {dk^+ \over k^{+}}\int \bigl({p^+ \over
\bar{p}^+} \bigr){1 \over {p^- - \bar{p}^-}}
\biggl[ 1 - {{m^2k^{+2} \over p^{+2}} \over {k_\perp^2 +{m^2k^{+2} \over
p^{+2}}}} \biggr] \Theta_{\Delta}(k) \;,
\end{displaymath}
which has a vanishing denominator at $q=0$ and therefore must be calculated
using the Heitler method
\cite{MUS91,HEIT54}. This is  easily seen to be free of IR divergences.

Thus,  one finally obtains
\begin{equation}
\Lambda^{\prime +}_{IR}(p,p)=-{{\lambda e^3} \over {2\pi^3}} \int {dk^+
\over k^{+}}\int d^2k_\perp
{\Theta_\Delta(k) \over {k_\perp^2 +{m^2k^{+2} \over p^{+2}}}}
\biggl[ 1 - {{m^2k^{+2} \over 2p^+} \over
{k_\perp^2 +{m^2k^{+2} \over p^{+2}}}} \biggr] \;.
\end{equation}

\begin{figure}
\caption{Vertex correction diagrams that contribute to $\Lambda^+(p,\bar{p})$.}
\label{reg}
\end{figure}

\begin{figure}
\caption{Contributions to the $O(e^2)$ vertex corrections due to emission and
absorption
of soft photons.}
\label{coh}
\end{figure}
\end{document}